\documentclass[twocolumn,nofootinbib,prd]{revtex4}
\usepackage{graphicx}
\usepackage{amsmath, amssymb}

\newcommand{\vev}[1]{\langle {#1} \rangle}
\newcommand{\lsim}{\lesssim}

\newcommand{\eq}[1]{Eq.~(\ref{#1})}
\newcommand{\beq}{\begin{equation}}
\newcommand{\eeq}{\end{equation}}
\newcommand{\bea}{\begin{eqnarray}}
\newcommand{\eea}{\end{eqnarray}}

\newcommand{\eps}{\varepsilon}

\newcommand{\mzd}{m_{Z_d}}

\newcommand{\diphot}{\gamma\gamma}

\newcommand{\gev}{{\rm GeV}}
\newcommand{\mev}{{\rm MeV}}

\begin{document}
\title{Dark Side of Higgs Diphoton Decays and Muon \boldmath{$g-2$}}
\author{Hooman Davoudiasl\footnote{email: hooman@bnl.gov}}
\author{Hye-Sung Lee\footnote{email: hlee@bnl.gov}}
\author{William J. Marciano\footnote{email: marciano@bnl.gov}}
\affiliation{Department of Physics, Brookhaven National Laboratory, Upton, NY 11973, USA}
\date{August 2012}

\begin{abstract}
We propose that the LHC hints for a Higgs diphoton excess and the muon $g-2$ ($g_\mu - 2$) discrepancy between theory and experiment may be related by vector-like ``leptons" charged under both $U(1)_Y$ hypercharge and a ``dark'' $U(1)_d$.
Quantum loops of such leptons can enhance the Higgs diphoton rate and also generically lead to $U(1)_Y$-$U(1)_d$ kinetic mixing.
The induced coupling of a light $U(1)_d$ gauge boson $Z_d$ to electric charge can naturally explain the measured $g_\mu - 2$.
We update $Z_d$ mass and coupling constraints based on comparison of the electron $g-2$ experiment and theory, and find that explaining $g_\mu - 2$ while satisfying other constraints requires $Z_d$ to have a mass $\sim 20 - 100 ~\mev$.
We predict new Higgs decay channels $\gamma Z_d$ and $Z_d Z_d$, with rates below the diphoton mode but potentially observable.
The boosted $Z_d \to e^+ e^-$ in these decays would mimic a promptly converted photon and could provide a fraction of the apparent diphoton excess.
More statistics or a closer inspection of extant data may reveal such events.
\end{abstract}
\maketitle

The discovery of a new Higgs-like state, at a mass of about $125
~\gev$, by the ATLAS \cite{ATLAS0712} and CMS \cite{CMS0712}
collaborations at the Large Hadron Collider (LHC) represents a
historical breakthrough in particle physics which is likely to
provide a major step toward understanding electroweak symmetry
breaking (EWSB). It remains to be seen whether this new state is the
long-sought Standard Model (SM) Higgs or some variant of it. While
addressing this important question requires more data, the current
results by both ATLAS and CMS hint, at a roughly $2 \sigma$ level
\cite{ATLAS0712,CMS0712}, that the diphoton branching fraction of
the new state, which we will henceforth refer to as the Higgs boson
$H$, seems to be a factor of $\sim 1.5-2$ larger than the SM
prediction \cite{Ellis:1975ap,Shifman:1979eb}. If that diphoton
excess is confirmed with more statistics it would be an important
clue for physics beyond the SM.

Besides the recent hints from the Higgs data, the current $3.6
\sigma$ discrepancy between the SM prediction and the measured value
of the muon $g-2$ \cite{Bennett:2006fi,PDG}, denoted by $g_\mu - 2$, is another
potential clue pointing to new physics \cite{Czarnecki:2001pv}. In
this work, we propose that the Higgs diphoton excess and the $g_\mu
- 2$ discrepancy can be naturally related through the introduction
of heavy new vector-like leptons charged under both $U(1)_Y$ as well as a
new $U(1)_d$ gauge symmetry with a relatively light ${\cal O} (20 -
100 ~\mev)$ $Z_d$ boson \cite{SUSYframework}.
(The lower bound of $20 ~\mev$ will be discussed later.)
We will refer to the
new quantum number as ``dark charge'' $Q_d$ since  $Z_d$ does not
have direct couplings to the ordinary ``visible'' SM sector.
However, our study needs not assume a specific connection with dark matter (DM) physics, although it is a possibility.
This issue will be briefly addressed.

To set the stage for our discussion, we first address the main
problem with trying to explain an excessive Higgs diphoton, $H \to
\diphot$, decay rate. The SM prediction arises from destructive
interference of $W^\pm$ loops (the dominant contribution) and a
smaller (relatively negative) top quark loop. Adding new heavy
charged chiral fermions leads to additional negative loop
contributions which further reduce the diphoton Higgs decay
amplitude. (We do not consider cases where many fermions are added
which change the overall sign and magnitude of the $H \diphot$
amplitude.) However, if the new fermions are vector-like doublets
and singlets, the existence of heavy gauge invariant masses combined
with mixing induced by the Higgs-singlet-doublet Yukawa couplings
can change the sign of the new fermion loop contribution and
actually enhance the Higgs diphoton branching ratio. Variants of
that possibility have been suggested by a number of authors
\cite{Dawson:2012di,Carena:2012xa,Bonne:2012im,An:2012vp,Joglekar:2012hb,ArkaniHamed:2012kq,Almeida:2012bq,Kearney:2012zi,Ajaib:2012eb}
who have discussed such scenarios in detail, including experimental constraints
on properties of new vector-like fermions \cite{SUSYdiphot}.

Here, we assume the above solution to the diphoton excess as our
starting point, but endow the new fermions with an additional
$U(1)_d$ gauge symmetry with a light gauge boson, $Z_d$ of mass
$m_{Z_d} \simeq 20 - 100 ~\mev$. To avoid changing the Higgs
production rate through gluon fusion, thereby affecting rates for
other final states, these fermions should not carry $SU(3)_C$ color
quantum numbers. Hence, we focus on new ``charged leptons.''

One-loop diagrams involving the new vector-like leptons can also
induce, via kinetic mixing, a suppressed coupling of $Z_d$ to
ordinarily charged particles. Such a gauge boson has recently been
invoked in generic discussions of DM particles and their
potential phenomenology \cite{DMzprime,Davoudiasl:2010am}. As we
shall see, the $Z_d$ provides a natural viable solution to the $g_\mu - 2$
discrepancy for a narrow range of $m_{Z_d}$ and leads to predictions for new Higgs decay modes, $H
\to \gamma Z_d$ and $Z_d Z_d$ at the LHC, which may occur at
observable rates or if not seen, used to constrain such models.

\begin{figure}[tb]
\begin{center}
\includegraphics[width=0.233\textwidth]{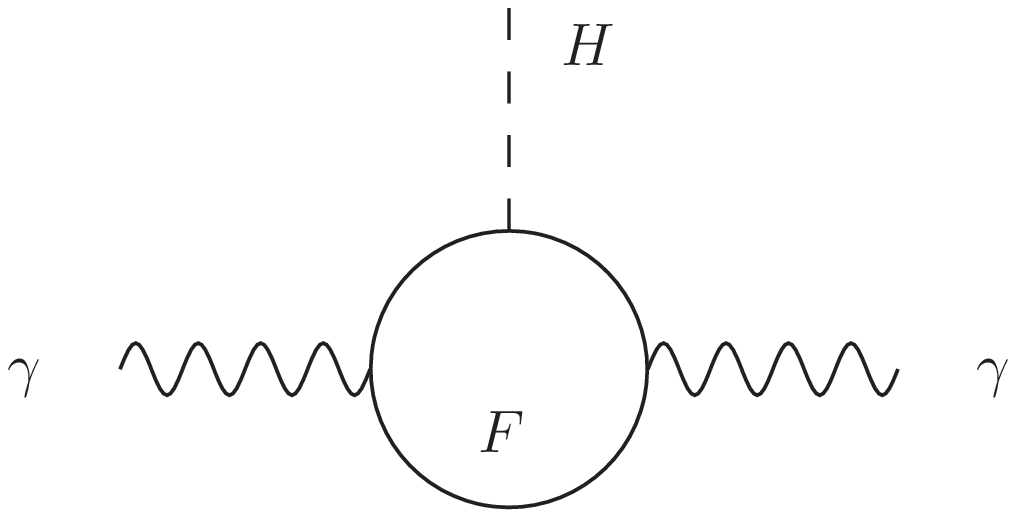} ~
\includegraphics[width=0.233\textwidth]{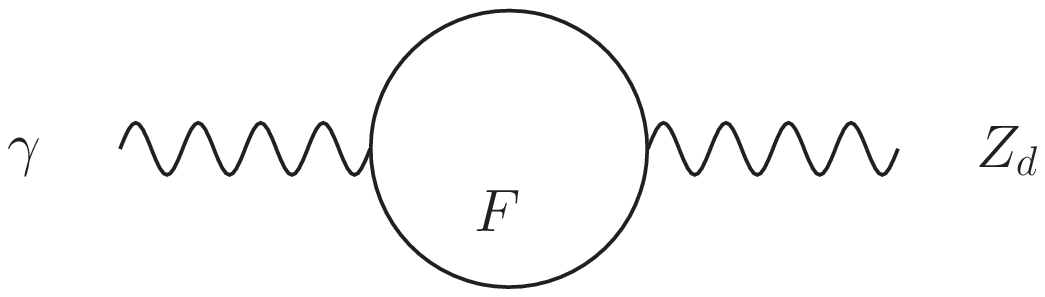}
\end{center}
\caption{New fermion ($F$) loop contribution to $H \to \diphot$ decay (left) and $\gamma-Z_d$ kinetic mixing (right).}
\label{fig:motivation}
\end{figure}

Vector-like heavy fermions can be found in
various new physics models including composite Higgs
\cite{Kaplan:1983sm} and supersymmetric $U(1)'$ models
\cite{Langacker:2008yv} motivated to address the $\mu$-problem
\cite{Kim:1983dt}. To avoid problems with new stable charged
particles, we consider charges which are multiples of the SM particle
charges. This would typically imply that the new particles carry
hypercharge (for the sake of minimality, we do not consider $SU(2)$
triplet fermions as they would require additional Higgs content).

Let us denote the ratio of the enhanced rate for $H\to \diphot$
compared to that in the SM by
\beq
R_{\diphot}\equiv
\frac{\Gamma(H\to \diphot)}{\Gamma(H\to \diphot)_{\rm SM}}.
\label{R2gam}
\eeq
Using the results of
Ref.~\cite{ArkaniHamed:2012kq}, the contribution of a new fermion
$F$ of electric charge $Q$ and mass $m_F$ (see
Fig.~\ref{fig:motivation}) to the above ratio is given by
\beq
R_{\diphot} = \left|1 + \frac{(4/3) Q^2}{A_{\rm SM}}\frac{\partial
\log m_F}{\partial \log v} \left(1 + \frac{7 m_H^2}{120
m_F^2}\right)\right|^2\,,
\label{RF}
\eeq where $A_{\rm SM} \simeq
-6.5$ stems from the SM amplitude for the decay, $v = \sqrt{2}
\vev{H} \simeq 246$~GeV, and $m_H \simeq 125$~GeV is the Higgs mass.
\eq{RF} implies that for $R_{\diphot}>1$ we need $\partial \log
m_F/\partial \log v < 0$, that is, the contribution of EWSB ($v\neq
0$) to the mass of $F$ must be negative.

We next describe a simple scenario which explains the diphoton excess.
We extend one of the examples proposed in Ref.~\cite{ArkaniHamed:2012kq}
to include dark $U(1)_d$ interactions.
The extended model contains vector-like ``leptons'', $\psi$ and
$\chi$ with $SU(3)_C \times SU(2)_L \times U(1)_Y \times U(1)_d$ charge assignments
\beq
(\psi,\psi^c)
\sim (1,2)_{\{\pm \frac{1}{2}, \pm 1\}}\quad ; \quad
(\chi,\chi^c)
\sim (1,1)_{\{\mp 1, \mp 1\}} ,
\label{charges}
\eeq
where upper (lower) signs are for $\psi$ ($\psi^c$), {\it etc}.
Here, electric charge $Q$ is related to hypercharge $Y$
and the third component of isospin $T_3$ by $Q = T_3 + Y$.
The mass matrix is obtained from the following Lagrangian:
\beq
- {\cal L}_m = m_\psi \psi \psi^c + m_\chi \chi \chi^c + y H \psi \chi +
y^c H^\dagger \psi^c \chi^c + {\rm h.c.}
\label{Lpsichi}
\eeq
After EWSB, we get two Dirac fermions $L_1$ and $L_2$
with electric charge $|Q| = 1$ and masses $m_1$ and $m_2$, where $m_1 + m_2 = m_\psi + m_\chi$, $m_2 > m_1$
and one neutral Dirac fermion $N$ of mass $m_\psi$ with $m_1 < m_\psi < m_2$.
All the new heavy leptons have pure vector couplings to gauge bosons; hence, no gauge anomalies are present.

One can show that the above field content results in an enhancement of the diphoton rate,
which to a good approximation is given by \cite{ArkaniHamed:2012kq}
\beq
R_{\diphot} \simeq \left|1 + 0.1 \, \frac{\Delta_v^2} {1 + \sqrt{\Delta_v^2 + \Delta_m^2}}\right|^2,
\label{R2gammodel}
\eeq
which is valid for $y\simeq y^c$, with
\beq
\Delta_v^2 \equiv \frac{2 \,y y^c v^2}{m_1^2} \;\; \text{and} \;\;
\Delta_m^2 \equiv \frac{(m_\psi - m_\chi)^2}{m_1^2}\,.
\label{Deltavm}
\eeq
The enhancement $R_{\diphot} \sim 1.5$ can be achieved for $y y^c \sim 1$ and $m_1 \sim m_H$,
assuming $\Delta_m^2 \ll \Delta_v^2$.

Without further assumptions, the model in Eqs.~(\ref{charges}) and
(\ref{Lpsichi}) will lead to stable charged particles since the lightest new vector-like lepton $L_1$ will be charged.
To avoid this unphysical situation, following Ref.~\cite{ArkaniHamed:2012kq}, we mention two approaches:

(I) One possibility is to allow very small mass mixing between the new charged heavy leptons $L_i$ ($i = 1, 2$), and ordinary SM charged leptons $\ell_j$ ($j = e, \mu, \tau$) which gives rise to $L_i \to \ell_j Z_d$ decays.
For us, this could be simply achieved if we introduce a ``dark'' Higgs field, $\phi_d$, with $Q_d = -1$ which allows small Yukawa interactions $G_{ij} \phi_d L_i^c \ell_j$ where $G_{ij}$ is a general Yukawa coupling matrix which connects the new heavy charged leptons with ordinary charged SM leptons.
The left-right and right-left $G_{ij}$ couplings will in general be different.
The $\phi_d$ is well motivated since it can also provide a mechanism for spontaneous symmetry breaking and lead to $m_{Z_d} \lsim 100 ~\mev$ if $\left< \phi_d \right> \sim 100 ~\mev$ ($m_{Z_d} \sim g_d \left< \phi_d \right>$).

The Yukawa mixing interaction will induce potentially dangerous flavor changing weak neutral current interactions for the $Z$, $H$ and $Z_d$ bosons.
Here, we focus on the $Z_d$ couplings.
They are important because the small $m_{Z_d}$ leads to interesting enhancements and are specific to the model we are considering.
The induced non-diagonal interaction $G_{ij} (m_{Z_d} / m_{L_i}) Z_d^\mu \bar \ell_j \gamma_\mu L_i$ appears to be highly suppressed by the $m_{Z_d} / m_{L_i} < 10^{-3}$ factor.
However, for the $Z_d$ longitudinal component (or Goldstone mode $s_d^0$), that factor is cancelled \cite{Davoudiasl:2012ag} and one finds the coupling $\sim G_{ij} s_d^0 \bar \ell_j L_i$ as required by the Goldstone boson equivalence theorem with different $G_{ij}$ for left and right handed $L_i$.
To avoid generating large chiral changing loop effects in quantities such as the electron and muon anomalous magnetic moments, lepton number flavor violating amplitudes for $\mu \to e \gamma$, $\tau \to \mu \gamma$, light lepton loop induced masses, etc, some of the $| G_{ij} |$ (particularly those involving $e$ or $\mu$) have to be quite small $\lsim 10^{-3}$.
However, even with such small couplings, the decay rates $\Gamma (L_i \to \ell_j Z_d) \sim | G_{ij} |^2 m_{L_i}$ are likely to provide very prompt signals, $L_i \to \ell_j Z_d$, at the LHC, where the light $Z_d$, which subsequently decays into $e^+ e^-$, can mimic a converted high energy photon.
(One expects $L_i \bar L_i$ pairs at colliders actually giving rise to di-$\ell_j Z_d$ events.)  We note that by 
assuming small mixing parameters ($\lsim 10^{-3}$), we can avoid conflict with precision 
measurements of $e$-$\mu$-$\tau$ universality such as those discussed in Ref.~\cite{Batell:2012mj}.
Of course, one might use lepton mixing effects to 
make further predictions or to accommodate all or part of the $g_\mu - 2$ discrepancy.
The constraints on this model and its phenomenology are potentially rich and interesting, but beyond the scope of this paper.
Here, we only suggest it as a means to avoid stable heavy charged leptons.

(II) Another possibility is to avoid mixing with the SM leptons and instead add the fields $(n, n^c) \sim (1,1)_{\{0,\mp1\}}$
(carrying only $Q_d \neq 0$), with the new expanded neutral fermion mass terms
\beq
m_n n\, n^c +
y_n H^\dagger \psi \, n + y_n^c H \psi^c n^c + {\rm h.c.},
\label{nmass}
\eeq
which together with \eq{Lpsichi}
result in two neutral Dirac fermions, $n_1$ and $n_2$, with masses $m_{n_2} > m_{n_1}$ and potentially $m_{n_1} < m_1$.
Given the mixing terms in \eq{nmass}, the new charged $L_1$ particle would decay into (possibly virtual) $W^\pm$ and $n_1$.

It may be tempting to think of the lightest new neutral state, $n_1$ as a stable relic DM candidate.
However, without further assumptions, this turns out to be not phenomenologically viable.
We will address this question and discuss potentially viable DM alternatives in the appendix.

\begin{figure}[tb]
\begin{center}
\includegraphics[width=0.4\textwidth]{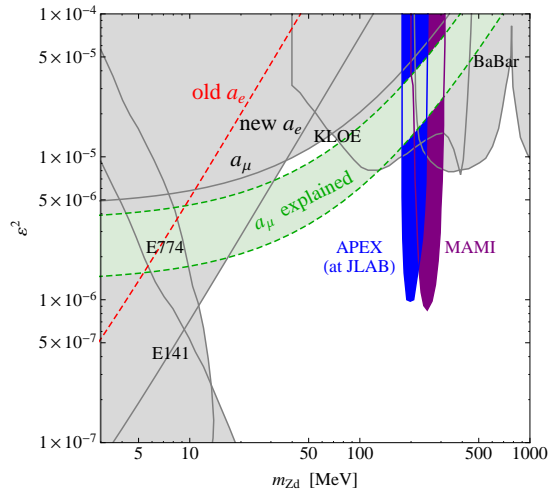}
\end{center}
\caption{Exclusion region in $m_{Z_d} - \eps^2$ space, updated from Refs.~\cite{McKeown:2011yj,Davoudiasl:2012qa} to include recent \cite{Hanneke:2008tm,Bouchendira:2010es,Aoyama:2012wj} $a_e$ results (``new $a_e$"); for comparison we also present the old bound (``old $a_e$").}
\label{fig:parameterSpace}
\end{figure}

We now turn to the $g_\mu - 2$ problem which has persisted over the last several years.
The discrepancy between the measured value and the SM prediction is about $3.6 \sigma$ \cite{Bennett:2006fi,PDG}:
\beq
\Delta a_\mu = a_\mu^{\rm exp} - a_\mu^{\rm SM} = 287 (80) \times 10^{-11}\,,
\label{Delamu}
\eeq
where $a_\mu \equiv (g_\mu - 2)/2$.
A simple explanation \cite{Zdgm2} of this difference postulates a new hidden $U(1)_d$ symmetry which kinetically mixes with $U(1)_Y$ by
\beq
{\cal L}_{\rm km} = \frac{1}{2} \frac{\eps}{\cos\theta_W} B_{\mu\nu} Z_d^{\mu\nu}\,,
\label{Lkm}
\eeq
where $\eps$ parametrizes the mixing, $\theta_W$ is the weak mixing angle,
and $X_{\mu\nu} = \partial_\mu X_\nu - \partial_\nu X_\mu$, $X=B,Z_d$, is a $U(1)$ field strength tensor.
Upon kinetic diagonalization, one finds that the massive dark boson $Z_d$
obtains an induced coupling  $e \,\eps Z_d^\mu J_\mu^{em}$, where $J_\mu^{em}$ is the
the electromagnetic current \cite{darkZ}.
This light boson is a target of active and planned searches at JLAB and MAMI in Mainz \cite{Bjorken:2009mm,Freytsis:2009bh,McKeown:2011yj,Merkel:2011ze,Abrahamyan:2011gv,DarkLight,Russell:2012zz}.
Early results from those experiments are illustrated as constraints in Fig.~\ref{fig:parameterSpace}.

One can show that the 1-loop contribution of the $Z_d$ to $a_\mu$ is given by \cite{g-2,Zdgm2}
\beq
a_\mu^{Z_d}  = \frac{\alpha}{2\pi} \eps^2 F_V\left(m_{Z_d} / m_\mu \right) \label{eq:5}
\eeq
with
$
F_V(x) \equiv \int_0^1 dz [2 z (1-z)^2]/[(1-z)^2 + x^2 z] \, ; F_V(0) = 1 .
$

In Fig.~\ref{fig:parameterSpace}, we give the current exclusion bounds on $\eps^2$ (adopted from Refs.~\cite{McKeown:2011yj,Davoudiasl:2012qa}).
There, we have updated the bounds coming from the recently improved electron anomalous magnetic moment comparison between experiment \cite{Hanneke:2008tm} and SM theory \cite{Bouchendira:2010es,Aoyama:2012wj}:
\beq
\Delta a_e = a_e^\text{exp} - a_e^\text{SM} = -1.06 (0.82) \times 10^{-12} .
\label{eq:ae}
\eeq
Because of the small momentum transfer in Rydberg measurements $Q^2 \ll m_{Z_d}^2$, the effect of a light $Z_d$ on the determination of $\alpha$ in Ref.~\cite{Bouchendira:2010es} is expected to be negligible for the $m_{Z_d}$ mass range considered.
That constraint implies at the $3 \sigma$ level $a_e^{Z_d} < 1.4 \times 10^{-12}$ (with $a_e^{Z_d}$ obtained from \eq{eq:5} with $m_\mu \to m_e$) which rules out a significant region which would otherwise provide a viable $\Delta a_\mu$ solution (i.e. $m_{Z_d} \lsim 20 ~\mev$, $\eps^2 \sim {\cal O}(2-5 \times 10^{-6})$.
Hence, one finds that the discrepancy in \eq{Delamu} can be explained for
\beq
20 ~\mev \lsim m_{Z_d} \lsim 100 ~\mev
\label{mZdrange}
\eeq
and
\beq
2 \times 10^{-6}\lsim \eps^2 \lsim 10^{-5} \quad
\label{eps2range}
\eeq
without conflict with current experimental bounds on $Z_d$.
We note that the exclusion region due to $a_e$ is somewhat enhanced because of the sign of \eq{eq:ae} which is opposite to that expected from $Z_d$.

Our new constraint rules out the (previously allowed) $10 ~\mev \lsim m_{Z_d} \lsim 20 ~\mev$ region of the ``$a_\mu$'' band in Fig.~\ref{fig:parameterSpace}.
That explicit part of the band is the focus of a proposed direct $Z_d$ search at VEPP-3 \cite{Wojtsekhowski:2012zq}; however, the experiment will also explore smaller $\eps^2$.

In a simple $U(1)_Y\times U(1)_d$ framework, $\eps$ is an arbitrary renormalized parameter set by experiment.
Normally, we expect $\eps \simeq e g_d / 8 \pi^2$ which for $g_d \simeq e$ gives $\eps \simeq 10^{-3}$.
That expectation would be natural, if either of the $U(1)$ symmetries at low energy descend from a non-abelian group in the ultraviolet sector, such that the high energy (bare) value of $\eps$ is zero.
In that case, finite kinetic mixing of the type in \eq{Lkm} can be naturally induced by loops of fermions which are charged under both $U(1)_Y$ and $U(1)_d$ \cite{Holdom:1985ag}.
The typical value of the kinetic mixing at low energies can then be estimated from a 1-loop diagram (see Fig.~\ref{fig:motivation}), which is roughly in the range indicated by \eq{eps2range}.
In fact, given an appropriate assignment of fermion charges, one can calculate a finite 1-loop result \cite{Holdom:1985ag}.
For example, a 2-generation extension of the reference model in \eq{charges} with opposite sign dark charges
$Q_d$ results in a finite and computable value
\beq
\eps = \frac{e\, Q_d \, g_d}{6 \pi^2}\log\left(\frac{m_1 m_4}{m_2 m_3}\right)\,,
\label{epsmodel}
\eeq
where $m_{3,4}$ correspond to the masses from the second generation of charged vector-like leptons.
We see that for typical values of parameters, $\eps$ in \eq{epsmodel} has the size needed to address the $g_\mu - 2$ anomaly.
For example, if the logarithm involving the masses is order unity, for $Q_d \sim 1$ and $g_d \sim e$ we find $\eps\sim 1.5 \times 10^{-3}$.
So, the model introduced in \eq{charges} to explain a $H \to \diphot$ excess can also accommodate the $g_\mu - 2$ discrepancy.

\begin{figure}[tb]
\begin{center}
\includegraphics[width=0.233\textwidth]{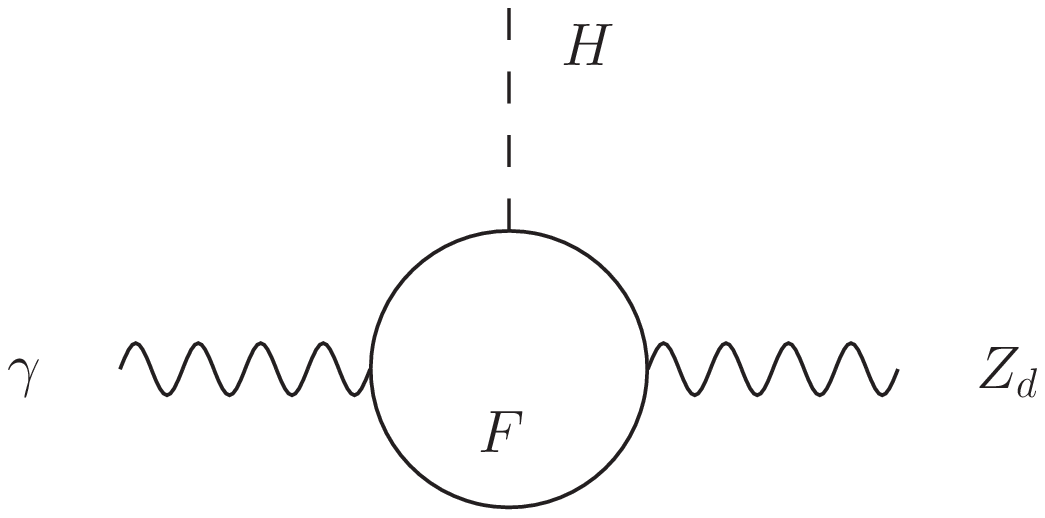} ~
\includegraphics[width=0.233\textwidth]{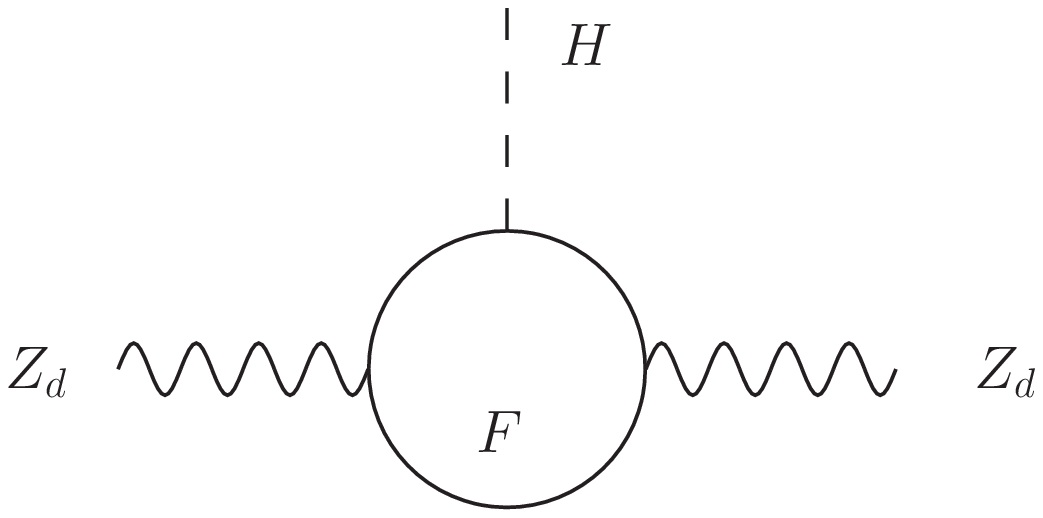}
\end{center}
\caption{New fermion ($F$) loop contribution to $H \to \gamma Z_d$ decay (left) and $H \to Z_d Z_d$ decay (right).
Both charged and neutral fermions should be included in the loop on the right.}
\label{fig:prediction}
\end{figure}

Our model leads to the interesting prediction that the 
Higgs has new decay channels $\gamma Z_d$ and $Z_d Z_d$ 
(Fig.~\ref{fig:prediction}) with rates somewhat smaller than that of the $\diphot$ mode.
To see this, note that Eqs.~(\ref{eps2range}) and (\ref{epsmodel}) imply we need $g_d \sim e$ to explain the measured $g_\mu - 2$.  Hence, we expect that the new decay modes in Fig.~\ref{fig:prediction} will have a similar amplitude as the extra contribution to $\diphot$ (Fig.~\ref{fig:motivation}).

To connect the Higgs diphoton excess and $g_\mu-2$, as proposed here, it is sufficient
to have a single fermion which carries both $U(1)_Y$ and $U(1)_d$ charges.  This minimal setup
can effectively emerge in our reference model in \eq{charges} if there is a modest hierarchy of masses and
the new Higgs decay amplitudes (Figs.~\ref{fig:motivation} and \ref{fig:prediction})
are dominated by the lightest charged state.
Under such a simplifying assumption, a rough estimate for the rate of the $H \to \gamma Z_d$
compared to the observed rate of $H \to \diphot$ (not the SM expectation) can be given in terms of $R_{\diphot}$ by
\beq
r_{\gamma Z_d} \equiv
\frac{\Gamma(H\to \gamma Z_d)}
{\Gamma(H\to \diphot)} \approx 2\left(1 - \frac{1}{\sqrt{R_{\diphot}}}\right)^2 \left(\frac{g_d}{e}\right)^2\,,
\label{GamgZd}
\eeq
where the factor of 2 accounts for the nonidentical final state particles, and the first set of parentheses factors out the new lepton contribution to the $H \to \diphot$ decay.
Similarly, the rate of the $H \to Z_d Z_d$ compared to $H \to \diphot$ is
\beq
r_{Z_d Z_d} \equiv
\frac{\Gamma(H\to Z_d Z_d)}{\Gamma(H\to \diphot)}
\approx \left(1 - \frac{1}{\sqrt{R_{\diphot}}}\right)^2 \left(\frac{g_d}{e}\right)^4\, .
\label{GamZdZd}
\eeq

\begin{figure}[tb]
\begin{center}
\includegraphics[width=0.4\textwidth]{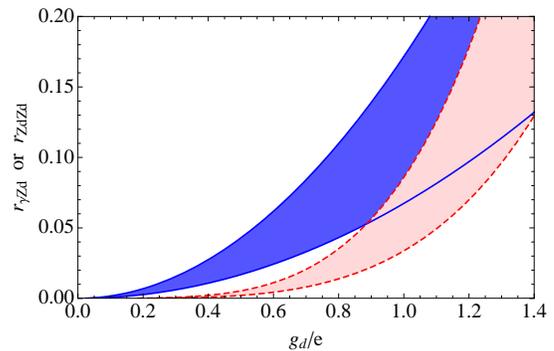}
\end{center}
\caption{$r_{\gamma Z_d}$ ($r_{Z_d Z_d}$) bounded by blue solid (red dashed) curves versus $g_d / e$, for $R_{\diphot} = 1.5$ (bottom) $-\ 2.0$ (top).}
\label{fig:BR}
\end{figure}

Based on our preceding discussion, let us take $g_d = e$ as a
typical value for our scenario. This would imply $r_{\gamma Z_d} \approx 0.07 -
0.17$ and $r_{Z_d Z_d} \approx 0.03 - 0.09$ for $R_{\diphot} = 1.5 - 2.0$ (see Fig.~\ref{fig:BR}).
We see that the rates for these new channels are expected to be well below that of $H\to \diphot$ but potentially within the reach of the LHC experiments. Note that for $\mzd$ in the range of \eq{mZdrange}, the $Z_d$ will mainly decay promptly into $e^+e^-$ \cite{Davoudiasl:2012ag},
i.e. within the beam pipe region.
However, since $\mzd\ll m_H$, the decay products will be boosted and highly collimated.
In general, we expect that the $Z_d$ in the final state will mimic a promptly converted photon ($\gamma \to e^+ e^-$) but with a small nonzero mass and production vertex near the beam rather than in the tracker.
Such properties could be used as a signal for these new decays.

For the $m_{Z_d} = 20 - 100 ~\mev$ range, the opening angle of order 
$0.002$ is well below the $0.02$ roughly required for separating the 
$e^+e^-$; so, except for the effect of the magnetic field in the detector, the $e^+e^-$ 
pair would be indistinguishable from a photon (see Ref.~\cite{Draper:2012xt} for some related discussions).
However, the magnetic field will lead to separated tracks for $e^-$ and $e^+$ in the tracker.  Under the assumption of a converted photon, these tracks would normally be fitted assuming zero invariant mass.
For our purposes, it would be useful to modify the fitting program to allow masses $20 - 100 ~\mev$.  
A careful examination of the available diphoton data, which is outside the scope
of this work and more appropriate for experimental scrutiny, could reveal or constrain the presence of 
$H \to \gamma Z_d$ or even $H \to Z_d Z_d$ events.

The above rough estimates of the relationships between
$\diphot$, $\gamma Z_d$ and $Z_d Z_d$ decay rates
can change if the underlying model deviates from
the simplifying assumptions used to derive Eqs.~\eqref{GamgZd} and \eqref{GamZdZd}.  For example,
in a 2-generation extension of the model,
the charged leptons carrying opposite dark charges
are likely to cancel partially and somewhat reduce the rate of $H \to \gamma Z_d$ (Fig.~\ref{fig:prediction}).
In this circumstance, our estimate of the $H \to \gamma Z_d$ should be viewed as an approximate upper bound.
In the case of the $H \to Z_d Z_d$ amplitude (Fig.~\ref{fig:prediction}), if the model is enlarged to include
the interactions in \eq{nmass}, the resulting neutral leptons
could either increase or reduce the rate for $H \to Z_d Z_d$.
In addition, a light $Z_d$ may be only a small part of the $\Delta a_\mu$ discrepancy, consistent with a much smaller $\eps^2$ below the ``$a_\mu$ explained'' band in Fig.~\ref{fig:parameterSpace}.
In that case, the $\gamma Z_d$ and $Z_d Z_d$ rates could be much reduced.

Overall, our estimates for the new $H$ decay rates are meant
to be suggestive and to stimulate experimental searches for those decays.  Definitive
predictions require more detailed studies using specific models and parameters.
We note that the search for $H \to \gamma Z_d$ and $H \to Z_d Z_d$
with $Z_d$ resembling a promptly converted photon in the LHC
experiments would be largely complementary to light ``dark boson'' searches such as those at JLAB and in Mainz \cite{Bjorken:2009mm,Freytsis:2009bh,McKeown:2011yj,Merkel:2011ze,Abrahamyan:2011gv,DarkLight,Russell:2012zz}.
Those programs will not only directly probe the $a_\mu$ explained region in Fig.~\ref{fig:parameterSpace} for $Z_d$ boson, but will also explore significant parts of parameter space outside that band.

``Dark'' decay modes of the Higgs may also arise through other mechanisms \cite{Gopalakrishna:2008dv,Baumgart:2009tn,Davoudiasl:2012ag}. 
For example, it is possible to have a $H \to Z_d Z_d$ mode from Higgs ($H$) and dark Higgs ($\phi_d$) mixing \cite{Gopalakrishna:2008dv},
and in the presence of mass mixing between $Z$ and $Z_d$, as studied in Ref.~\cite{Davoudiasl:2012ag},
a new Higgs decay mode $H\to Z Z_d$ would also be possible.  The latter channel could mimic $Z \gamma$
with a promptly converted photon if $Z_d$ is sufficiently light.
However, the predictions in those cases are more arbitrary; whereas the connection between the Higgs diphoton rate and $g_\mu - 2$ in our model allows us to make more quantitative estimates for the dark decay rates of the Higgs.
We also note that our loop induced $\gamma Z_d$ and $Z_d Z_d$ $H$ decays involve primarily transverse $Z_d$ bosons while the $Z_d Z_d$ and $Z Z_d$ decays in Refs.~\cite{Gopalakrishna:2008dv,Davoudiasl:2012ag} are dominated by longitudinal $Z_d$.

In this paper, we have discussed a possible link between the
reported excess of the Higgs to diphoton decay at the LHC
experiments and $g_\mu - 2$ via heavy new vector-like leptons and a light dark gauge
boson. A gauge boson of mass $\sim 20 - 100 ~\mev$ with small
induced coupling to the SM particles is well motivated as a rather simple
explanation of the $3.6 \sigma$ deviation of $g_\mu - 2$ from
the SM. The required coupling of the $Z_d$ to the SM fermions is
naturally obtained when it arises from loops of charged extra
fermions that couple to both the SM $U(1)_Y$ and a dark sector
$U(1)_d$ with similar size couplings. This scenario yields an
additional contribution to $H \to \diphot$ through a loop of the
charged extra fermions, which is consistent with the recent $2
\sigma$ level deviation at the LHC experiments.
The Higgs boson can also decay into $\gamma Z_d$ and $Z_d Z_d$,
with the light $Z_d$ bosons looking like
promptly converted photons in the ATLAS and CMS detectors.
Such a connection implies a few definite predictions in high energy experiments at the
LHC and complementary low energy $Z_d$ searches at JLAB and in Mainz.

\vspace{3mm}
Acknowledgments: This work was supported in part by the United States Department of Energy under Grant No. DE-AC02-98CH10886.
WM acknowledges partial support as a Fellow in the Gutenberg Research College.
We are grateful to I. Lewis for useful discussions.

\vspace{3mm}
Note Added:
After this paper was posted and submitted for publication, a preprint \cite{Endo:2012hp} appeared which reached similar conclusions regarding the updated exclusion region due to new $a_e$ results in Refs.~\cite{Bouchendira:2010es,Aoyama:2012wj}.
The authors of Ref.~\cite{Endo:2012hp} also gave a detailed analysis of 
the effect of a $Z_d$ on $\alpha$ explicitly showing that for the $m_{Z_d}$ 
range considered, it has a negligible effect.

\appendix
\section{Relation to Dark Matter}
As mentioned before, the lightest neutral Dirac lepton in our simple framework,
as presented above, is not a good DM candidate.  Starting from
our basic model in \eq{charges}, let us examine solutions (I) and (II), discussed earlier,
for avoiding a stable charged state $L_1$.
In case (I), a small degree of mass mixing with the SM leptons would not alter the
spectrum of the  model significantly, and we still expect the neutral state $N$ to be more massive than $L_1$ and thus allow the decay $N \to L_1 + (\text{virtual})~ W$.  Hence,
$N$ cannot be a long-lived DM candidate.

In case (II), the neutral particle $n_1$ is the lightest new state
and could be stable.  However, it carries dark charge and can hence elastically 
scatter from protons, through light $Z_d$ exchanges, at too high a
rate. For the parameter space of interest here, we would typically
expect the scattering cross section of $n_1$ from a nucleon to
be $\sim 10^{-33}$~cm$^2$ (see, for example, the last paper in
Ref.~\cite{DMzprime}) which, for masses $\sim 100 ~\gev$, is ruled
out by about 11 orders of magnitude from direct DM search
constraints \cite{:2012nq}.

One could arrange for the early universe annihilation cross section of $n_1$ to be
large, so that its relic density is very suppressed and it is not a primary component of DM.
Alternatively, one
could instead take a positive approach and extend our framework to allow for the presence
of a DM candidate. For example, let us postulate the $U(1)_d$ breaking
scalar $\phi_d$, with $\vev{\phi_d}\sim 100$~MeV, which was
invoked earlier in our discussion of case (I). We also add a singlet
vector-like lepton $\xi$ without any gauge charges and assume that
all of the new fermions are odd under a ${\mathbb Z}_2$ parity to
forbid DM decay. With $\xi$ in the spectrum, we can write down an
interaction $\lambda\, \phi_d \, n\, \xi$, with $\lambda \sim 1$.  If
$m_{n_1}> m_\xi$, the above interaction would lead to $n_1\to \phi_d
\, \xi$ which will be prompt, and all $n_1$ particles will decay into
$\xi$ and $\phi_d$. The scalar $\phi_d$ will eventually decay into SM
states. However, $\xi$ would be a viable, stable DM candidate. Note that due to the very 
small $n\,\xi$ mixing induced by $\vev{\phi_d}\neq 0$, the
interactions of $\xi$ with $Z_d$ could be quite suppressed, and hence
one would avoid the stringent bounds from direct detection.

The relic density of $\xi$ is set by $\xi\, \xi\to \phi_d \phi_d$,
through $t$-channel exchange of neutral states, which for weak scale
masses could be of the correct thermal relic size.
The $n\, \xi$ mixing is of order $\vev{\phi_d}/m_n \sim 10^{-3}$,
assuming $m_\xi\sim 100$~GeV, typical of the particles considered
here.  Hence, the direct detection cross section via $Z_d$ exchange
with protons could be suppressed by $\sim 10^{-12}$, which is just
below the current sensitivities \cite{:2012nq}.

An alternative possibility entails adding a lepton number violating
interaction to our model which splits the neutral Dirac lepton $n_1$
and its antiparticle partner into two nondegenerate Majorana states.
Such states have zero dark charge and will not scatter elastically
(at least at leading order) off ordinary matter via $Z_d$ exchange \cite{TuckerSmith:2004jv}.
There will be off-diagonal $Z_d$ couplings between the two Majorana states which allow inelastic
scattering, but for the lighter state, that can be kinematically suppressed.
A detailed evaluation of this scenario is beyond the scope of our study.



\end{document}